\documentclass[12pt]{iopart}

\usepackage{color}
\usepackage{subfig}
\usepackage{graphicx}
\usepackage{bm}
 \usepackage{ulem}
 \usepackage{amssymb}
\usepackage[colorlinks=true,linkcolor=green,citecolor=blue]{hyperref}
\usepackage{cite} 
\renewcommand{\thetable}{\arabic{table}}
\renewcommand{\thefigure}{\arabic{figure}}
\usepackage{CJKutf8}

\begin{document}
\begin{CJK*}{UTF8}{gbsn}

\title[]{Viscous effects on plasmoid formation from nonlinear resistive
tearing growth in a Harris sheet}
\author{Nisar AHMAD$^1$, Ping ZHU(朱平)$^{2,3\ast}$, Ahmad ALI$^4$, and Shiyong ZENG(曾市勇)$^5$}

\address{\small $^1$ CAS Key Laboratory of Geospace Environment and Department of Engineering and Applied Physics, University of Science and Technology of China, Hefei 230026, People's Republic of China}
\address{$^2$ International Joint Research Laboratory of Magnetic Confinement Fusion and Plasma Physics, State Key Laboratory of Advanced Electromagnetic Engineering and Technology, School of Electrical and Electronic Engineering, Huazhong University of Science and Technology, Wuhan, 430074, People's Republic of China}
\address{$^3$ Department of Engineering Physics, University of Wisconsin-Madison, Madison, Wisconsin 53706, USA}
\address{$^4$ Pakistan Tokamak Plasma Research Institute, Islamabad 3329, Pakistan}
\address {$^5$ Department of Plasma Physics and Fusion Engineering, University of Science and Technology of China, Hefei 230026, People's Republic of China}
\ead{zhup@hust.edu.cn}
\vspace{10pt}

\begin{abstract}\\
In this study, the evolution of a highly unstable ${m = 1}$ resistive tearing mode, leading to plasmoid formation in a Harris sheet is studied in the framework of full MHD model using the NIMROD simulation. Following the initial nonlinear growth of the primary $m = 1$ island, the X-point develops into a secondary elongated current sheet that eventually breaks into plasmoids. Two distinctive viscous regimes are found for the plasmoid formation and saturation. In the low viscosity regime (i.e. $P_r \lesssim 1$), the plasmoid width increases sharply with viscosity, whereas in the viscosity dominant regime (i.e. ${P_r \gtrsim 1}$ ), the plasmoid size gradually decreases with viscosity. Such a finding quantifies the role of viscosity in modulating the plasmoid formation process through its effects on the plasma flow and the reconnection itself. 

\vspace{2pc}
\newline
(Some Figures may appear in colour only in the online journal)
\end{abstract}


\section{Introduction}

Plasmoid instability (PI) is known to develop on the elongated current sheet formed during the
externally driven Sweet-Parker (SP) reconnection, or from the intrinsically growing nonlinear kink or tearing mode. In general, when the aspect ratio of the elongated current sheet becomes sufficiently large, unstable secondary tearing can lead to the formation of plasmoids \cite{MALARA1992, LOUREIRO2005}. The problem of the transition from the laminar reconnection during the early nonlinear stage, to the subsequent highly unstable one, characterized by sporadic production of plasmoids inside the sheet itself, with faster average reconnection rates, has been addressed by a number of past numerical and theoretical studies \cite{Lapenta2008, Bhattacharjee2009}, in the context of PIs following the externally driven SP reconnection \cite{LOUREIRO2007, Huang2010, Huang2013, Huang2011, LOUREIRO2013}, or the intrinsically nonlinear tearing mode \cite{LOUREIRO2005, ALI2014, Uzdensky2016, LOUREIRO2012}, on the scaling and dynamics of plasmoid formation with different Lundquist numbers.

Previous studies have found the critical roles of plasma flow in the processes of reconnection in general and plasmoid formation in particular \cite{DOBROTT1977, BULANOV1978, BULANOV1979, BISKAMP1986, TENERANI2015b,ALI2015}. Whereas the plasma outflow is stabilizing on the primary tearing mode or reconnection process \cite{DOBROTT1977, BULANOV1978, BULANOV1979, BISKAMP1986}, the effects of plasma flow itself, including both inflow and outflow, may contribute to the initial onset of plasmoid instability \cite{LOUREIRO2007}. The plasma viscosity can affect the properties and topologies of plasma flow close to the thin current sheet as well as the reconnection rate. Because of the narrowness of the current sheet, viscosity can influence the non-linear regime. In fact, viscosity increases the possibility of local changes in the flow topology. The robustness of the flow cells around the sheet might be weakened or even unstable due to the existence of strong flow gradients in the current sheet region \cite{Takeda2008}. Finite viscosity inserts dissipation to the flow patterns that in turn interact with the island evolution and reconnection \cite{Takeda2008}. Thus the plasma viscosity, both collisional and collisionless, is expected to be one of the key parameters that determine the onset and saturation conditions for the plasmoid instability. 

The effects of viscosity on linear and nonlinear resistive tearing mode as well as plasmoid instability have been studied by many \cite{LOUREIRO2013, ALI2014,BONDESON1984, PORCELLI1987, Ofman1991, Takeda2008, GRASSO2008, MILITELLO2011, TENERANI2015a, Betar2020, Comisso2016, Comisso2017}. In this paper we focus on exploring the impact of viscosity on the onset and saturation of plasmoid instability. Most of the past studies on visco-resistive tearing and kink modes were made in a 2D reduced MHD model, however in this study, we use the complete resistive MHD equations implemented in NIMROD code \cite{Sovinec2004}. Both the onset and the dynamics of plasmoid differ greatly from those found in previous reduced MHD simulations \cite{ALI2015,Takeda2008}.

The rest of the paper is organized as follows. In section 2, we briefly describe our simulation model. In section 3, both linear and nonlinear simulation results are reported. At the end in section 4, summary and discussion are presented.

\section{Simulation model and equilibrium} 

Our simulations are based on the single-fluid full MHD model implemented in the NIMROD (Non-Ideal Magnetohydrodynamics with Rotation, Open Discussion) code \cite{Sovinec2004}.

\begin{equation}
  \label{eq=one}
\frac{\partial \rho}{\partial t}+\nabla \cdot(\rho \mathbf{v})=0
\end{equation}

\begin{equation}
  \label{eq=one}
\rho\left(\frac{\partial}{\partial t}+\mathbf{v} \cdot \nabla\right) \mathbf{v}=\mathbf{J} \times \mathbf{B}-\nabla p+\rho \nu \nabla^{2} \mathbf{v}
\end{equation}

\begin{equation}
  \label{eq=one}
\frac{N}{\gamma-1}\left(\frac{\partial}{\partial t}+\mathbf{v} \cdot \nabla\right) \mathbf{T}=-{\frac{p}{2}} \nabla \cdot \mathbf{v}-\nabla \cdot \mathbf{q}
\end{equation}

\begin{equation}
  \label{eq=one}
\frac{\partial \mathbf{B}}{\partial t}=\nabla \times(\mathbf{v} \times \mathbf{B}-\eta\mathbf{J})
\end{equation}

\begin{equation}
  \label{eq=one}
\nabla \times \mathbf{B}=-\mu_{0} \mathbf{J}
\end{equation}

\begin{equation}
  \label{eq=one}
{\bf{q}}=-N\left(\chi_{\parallel} \nabla_{\parallel} {T}+\chi_{\perp} \nabla_{\perp} {T}\right)
\end{equation}
where ${\rho,N,p,\mathbf{J},\mathbf{v},\mathbf{B},{\bf{q}},\eta,\gamma,\nu,\chi_\parallel}$ and ${\chi_\perp}$ are the plasma mass density, number density, pressure,  current density, velocity, magnetic field, heat flux, resistivity, specific heat ratio, viscosity, parallel and perpendicular thermal conductivity, respectively. The Lundquist number ${S = \tau_R/\tau_A}$, where ${\tau_R= \mu_0a^2/\eta}$ is the resistive time and ${\tau_A= a/v_A}$ is the Alfvénic time (${a}$ represents the current sheet width), the Alfvén speed ${v_A=B_0 / \sqrt{\mu_0 \rho}}$ (${B_0}$ is the magnitude of magnetic field at the edge of the current sheet), and ${P_r = \nu/\eta}$ is the Prandtl number. The Harris current sheet model is adopted for the equilibrium magnetic field \cite{Harris196}

\begin{equation}
  \label{eq=one}
\mathbf{B_0}{(x)} = B_0 \mathrm{tanh} \left(\frac{x}{a}\right)\hat{\mathbf{y}}
\end{equation}
The corresponding pressure profile from the static MHD force balance is determined as

\begin{equation}
  \label{eq=one}
{P_0}{(x)} = \frac{B_0^2}{2\mu _0} \mathrm{sech^2} \left(\frac{x}{a}\right)
\end{equation}



The resistive MHD equations (1)–(6) are numerically solved in a rectangular domain ${[-L_x, L_x]\times{[-L_y, L_y]}}$. The periodic boundary conditions are imposed at the ${y}$-boundaries, and the solid, perfect conducting walls are assumed at the ${x}$-boundaries. For the Harris current sheet ${\Delta^{\prime}a = 2[(ka)^{-1} - ka]}$ \cite{Tenerani2016}, so that the unstable modes have wave vector satisfying ${ka < 1}$. Here ${\Delta^{\prime}}$ is the discontinuity of logarithmic derivative of the outer flux function when approaching the singular layer at ${x = 0}$, which is a measure of the free energy of the system. In our simulations ${\Delta^{\prime} = 49.66}$, ${k}$ is the mode wave number along ${y}$, and ${L_y = 2\pi m/k}$, with ${m}$ being the mode number. Simulations are performed for a uniform plasma resistivity ${\eta = 2.8 \times 10^{-4}}$ and a wide range of viscosity ( ${P_r}$ = 0.33 to 10), and the equilibrium plasma number density ${n = 3\times 10^{19}m^{-3}}$ .

\begin{figure}[htbp]
\centering
\begin{minipage}{0.7\textwidth}
\includegraphics[width=1.0\textwidth]{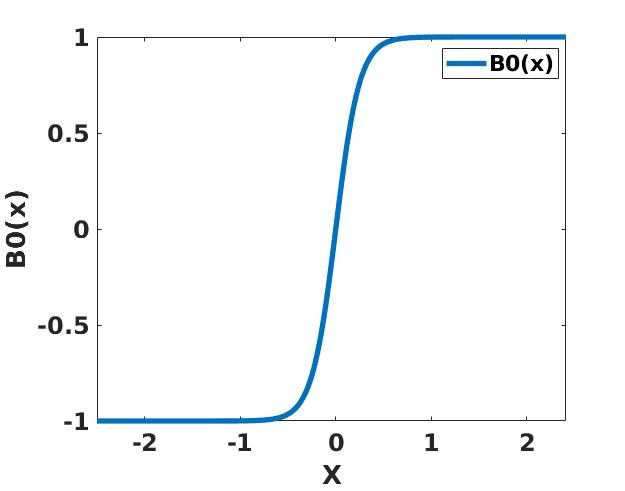}
\put(-270,235){\textbf{(a)}}
\end{minipage}
\begin{minipage}{0.7\textwidth}
\includegraphics[width=1.0\textwidth]{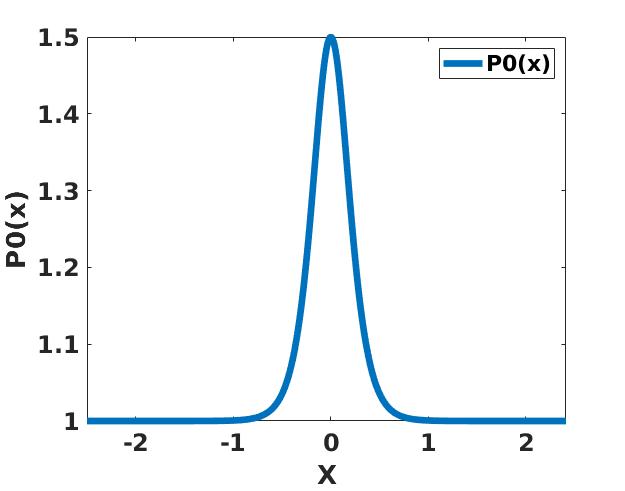}
\put(-270,235){\textbf{(b)}}
\end{minipage}
\caption{(a) Harris sheet equilibrium magnetic field and (b) pressure profiles.}
\end{figure}

\begin{figure}[htbp]
\centering
\begin{minipage}{0.7\textwidth}
\includegraphics[width=1.0\textwidth]{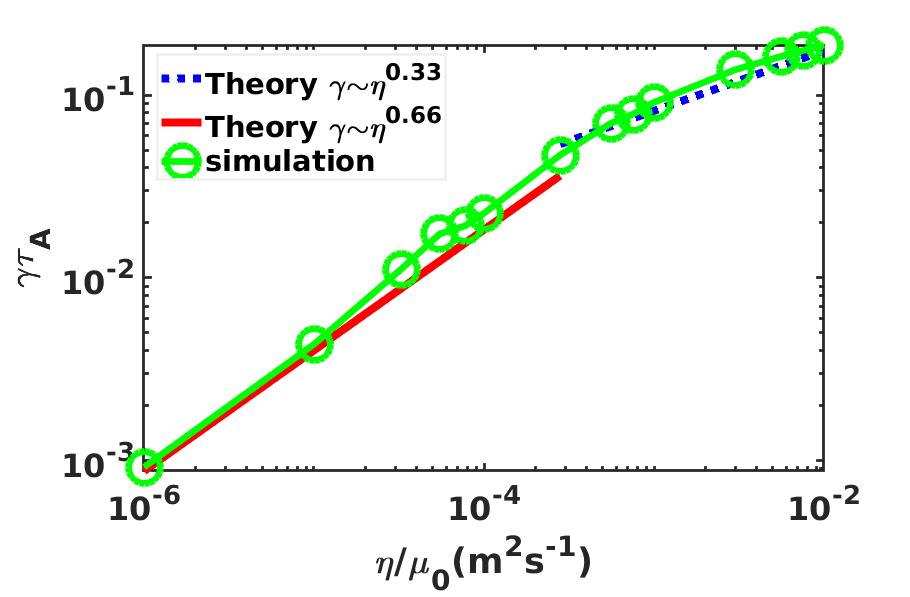}
\put(-270,210){\textbf{(a)}}
\end{minipage}
\begin{minipage}{0.7\textwidth}
\includegraphics[width=1.0\textwidth]{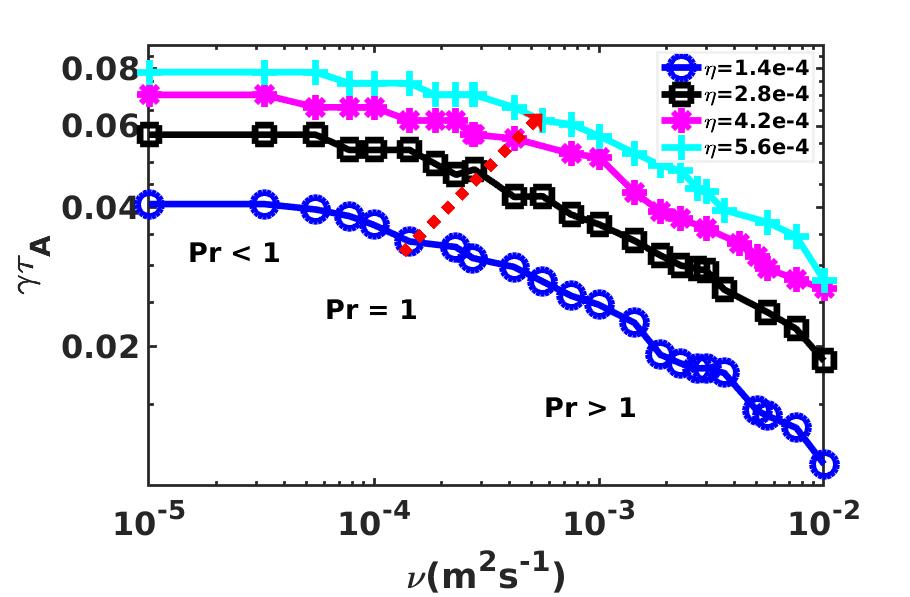}
\put(-270,210){\textbf{(b)}}
\end{minipage}
\caption{(a) Linear growth rates as functions of the resistivity for the fixed value of ${\Delta^\prime=49.66}$. The blue and red lines, represent the theoretical scaling of the linear growth rate for low (${\gamma \sim \eta ^{0.33}}$) and large (${\gamma \sim \eta ^{0.66}}$) Prandtl number regimes respectively, whereas, the green line represents the simulation results. (b) Linear growth rates as functions of the viscosity for the fixed value of instability parameter ${\Delta^\prime = 49.66}$ and various values of resistivity ${\eta}$.}
\end{figure}

\section{Simulation results}

\subsection{Linear scaling}

The plasmoid instability tends to develop from the primary tearing growth in the large ${\Delta^{\prime}}$ regime \cite{LOUREIRO2005}. One such case, ${\Delta^{\prime} = 49.66}$ is examined first in simulations for its linear scaling in comparison with theory. The linear growth rate of the ${m = 1}$ resistive tearing mode obtained from our NIMROD simulations scales with the resistivity ${\gamma}$ as ${\gamma \sim \eta^{0.30}}$, which is close to the resistive tearing scaling of ${\gamma \sim \eta^{0.33}}$ in the large ${\Delta^{\prime}}$ regime previously derived in theory \cite{Coppi1976, PORCELLI1987} (Figure 2a). Viscosity in general introduces dissipation that reduce the linear growth of resistive tearing mode. This viscous dissipation is stronger in the ${P_r > 1}$ regime, where the growth rate ${\gamma}$ of the ${m = 1}$ resistive tearing scales with the viscosity as ${\gamma \sim \nu^{-0.33}}$ , similar to the viscosity scaling obtained in previous reduced MHD simulations \cite{Betar2020} and theory \cite{PORCELLI1987} (Figure 2b).



\subsection{Nonlinear results}

\subsubsection{Critical $\Delta^{\prime}$ for the X-point collapse and plasmoid instability}~\\ 
Our nonlinear simulations find that the onset of secondary tearing instability and plasmoid formation occur only when the ${\Delta^{\prime}}$ is above a certain threshold value. In our nonlinear simulation, we mostly employ ${64 \times 48}$ 2D ﬁnite elements with a polynomial degree of 5, which ensures the numerical convergence (Figure 3). 

\begin{figure}[htbp]
\includegraphics[width=1\linewidth]{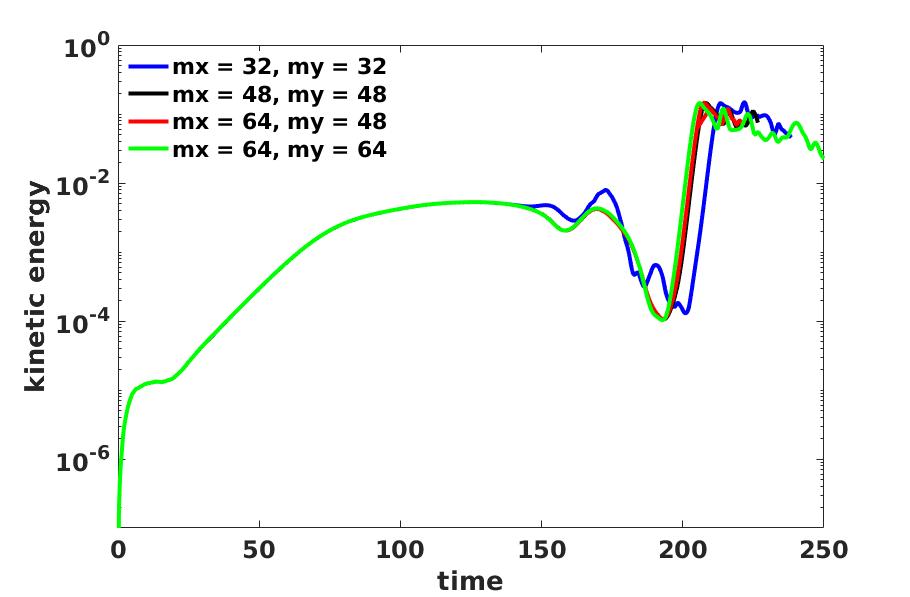}
\caption{Kinetic energy evolution for different numerical resolutions at ${\Delta^\prime = 49.66}$.}
\end{figure}

\begin{figure}[htbp]
\includegraphics[width=1\linewidth]{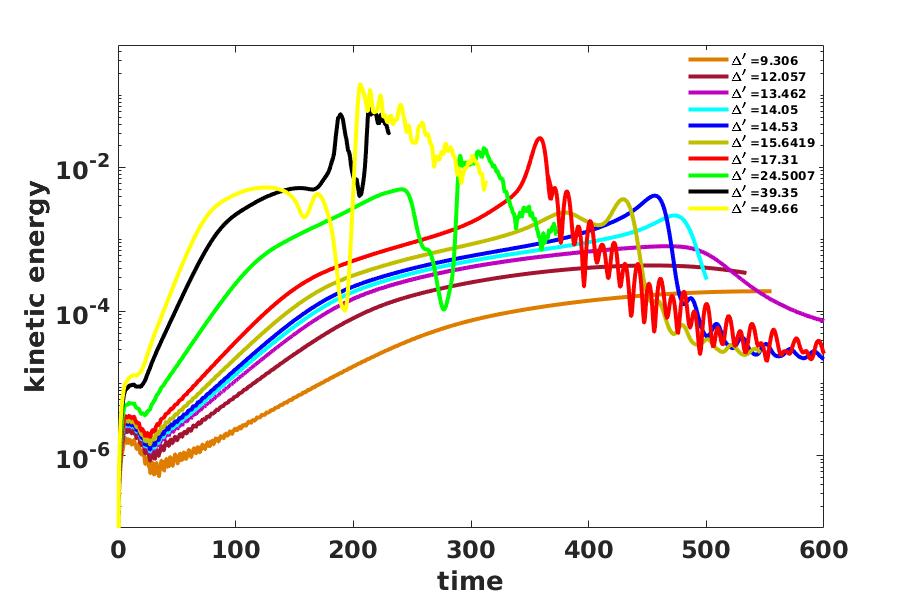}
\caption{Kinetic energy evolution for $\Delta^\prime$ = 9.306, 12.057, 13.462, 14.05, 14.53, 15.6419, 17.31, 24.5007, 39.35 and 49.66.}
\label{fig:10}
\end{figure}

\begin{figure}[htbp]
\centering
\begin{minipage}{0.55\textwidth}
\includegraphics[width=1.0\textwidth]{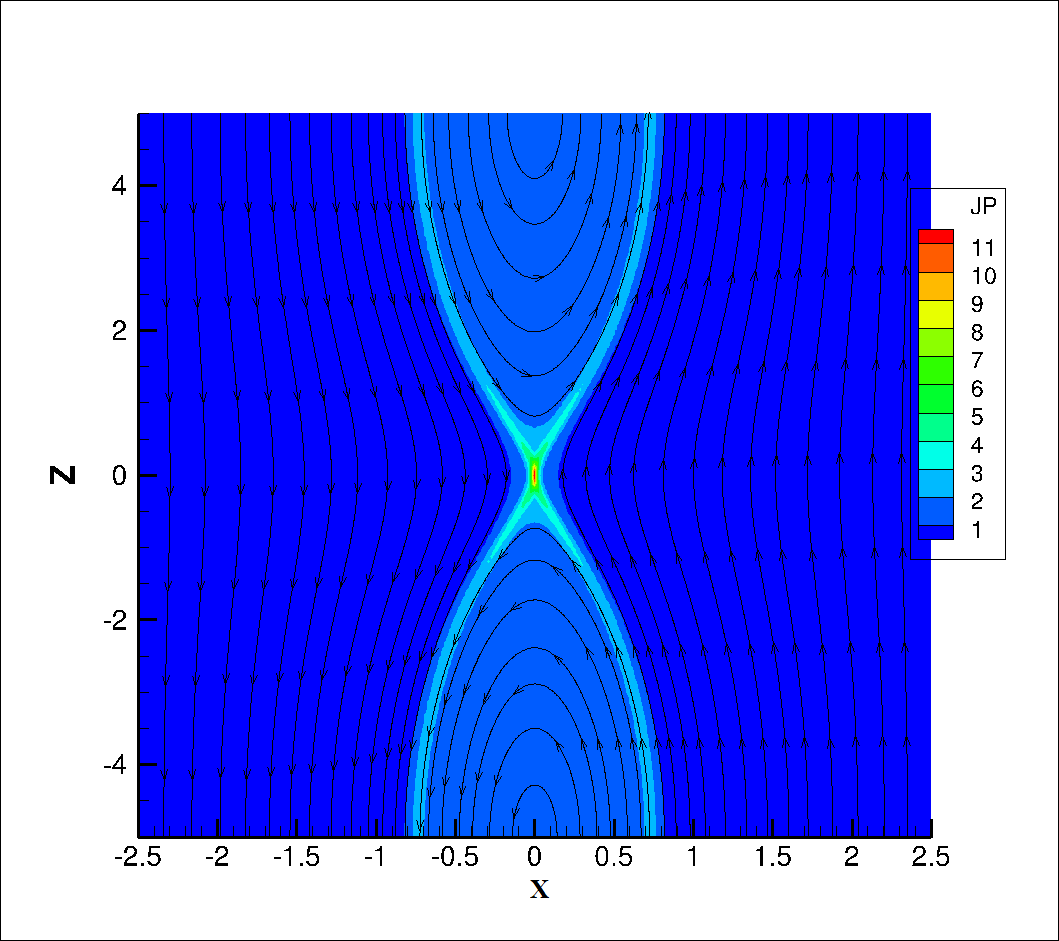}
\put(-250,225){\textbf{(a)}}
\end{minipage}
\begin{minipage}{0.55\textwidth}
\includegraphics[width=1.0\textwidth]{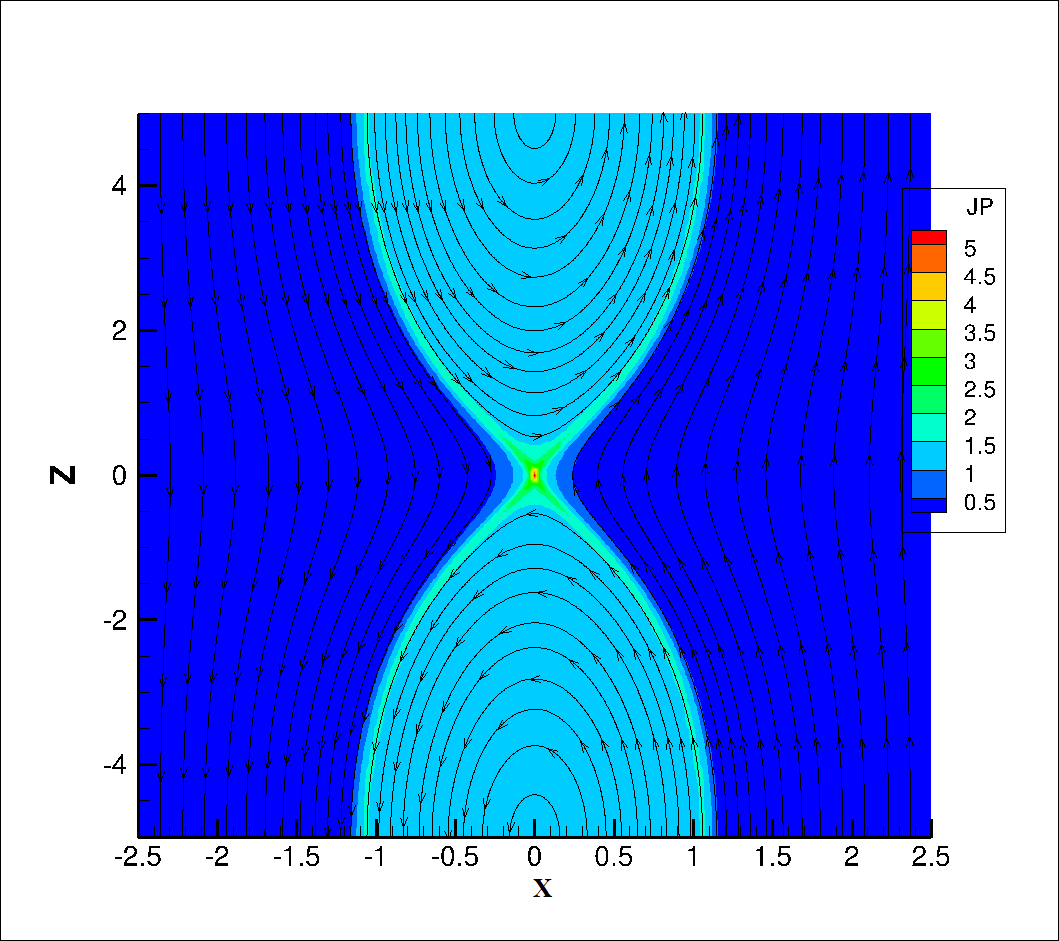}
\put(-250,225){\textbf{(b)}}
\end{minipage}
\caption{2D contours of the current density in z-direction and the 2D magnetic field lines at (a) time = 450 sec and (b) time = 570 sec for ${\Delta^\prime = 13.462}$.}
\end{figure}

The evolution of kinetic energy reaches its maximum sooner as we increase the value of ${\Delta^{\prime}}$ (Figure 4). The minimum value of ${\Delta^{\prime}}$ at which the X-point evolves into the Y-type is ${14.04}$. For ${\Delta^{\prime} = 13.42}$, the current sheet remains the shape of X-point over the entire time (Figure 5). As the value of ${\Delta^{\prime}}$ increases to above 14.05, the X-point evolves into a Y-type current sheet as shown in Figures 6 and 7. Waelbroeck \cite{Waelbroeck1993} first predicted the criterion for the collapse of X-point into Y-type current sheet to be ${W > W_c \approx 25/\Delta^{\prime}}$, where ${W}$ represents the width of primary island and ${W_c}$ represents the critical width at which X-point collapses. This conversion of X-point into Y-type current sheet is termed as the secondary instability (SI)\cite{Waelbroeck1993}. The subsequent collapse of the Y-type current sheet into plasmoids is known as the plasmoid instability (PI) \cite{Biskamp2000}.

For ${\eta = 2.8 \times 10^{-4}}$ and ${P_r = 1}$ the critical ${\Delta^{\prime}}$ value for the onset of plasmoid instability (PI) is ${\Delta^{\prime} = 17.31}$, which is in agreement with the previous reduced MHD simulation \cite{ALI2015}. In this case, at time 200 Sec, Figure 8(a), an X-point is formed as shown. At time 340 Sec, Figure 8(b), the Y-type current sheet develops during the nonlinear stage \cite{Lapenta2008, Huang2010, Uzdensky2010}. At time 360 Sec, Figure 8(c) the current density reaches maximum and the current sheet becomes more stretched and thinner. Finally, at time 365 Sec, Figure 8(d), formation of plasmoid takes place.

\begin{figure}[htbp]
\centering
\begin{minipage}{0.55\textwidth}
\includegraphics[width=1.0\textwidth]{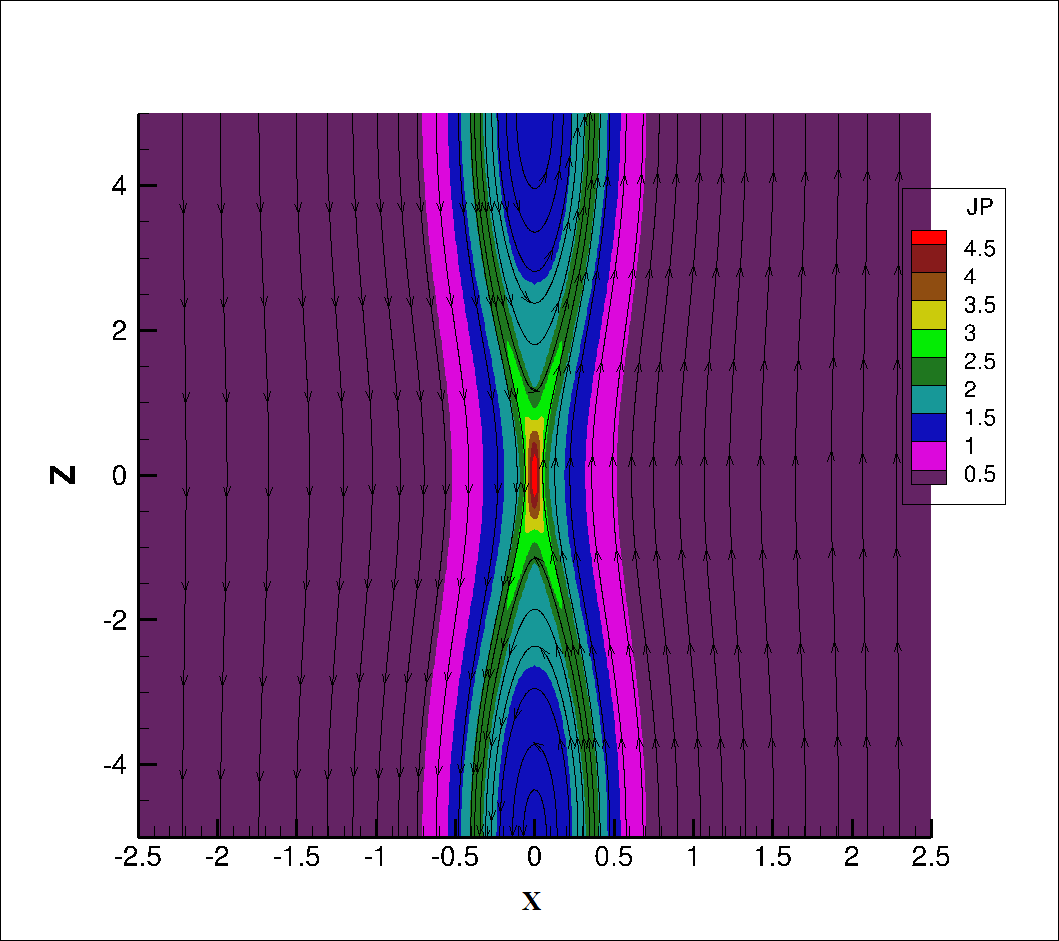}
\put(-250,225){\textbf{(a)}}
\end{minipage}
\begin{minipage}{0.55\textwidth}
\includegraphics[width=1.0\textwidth]{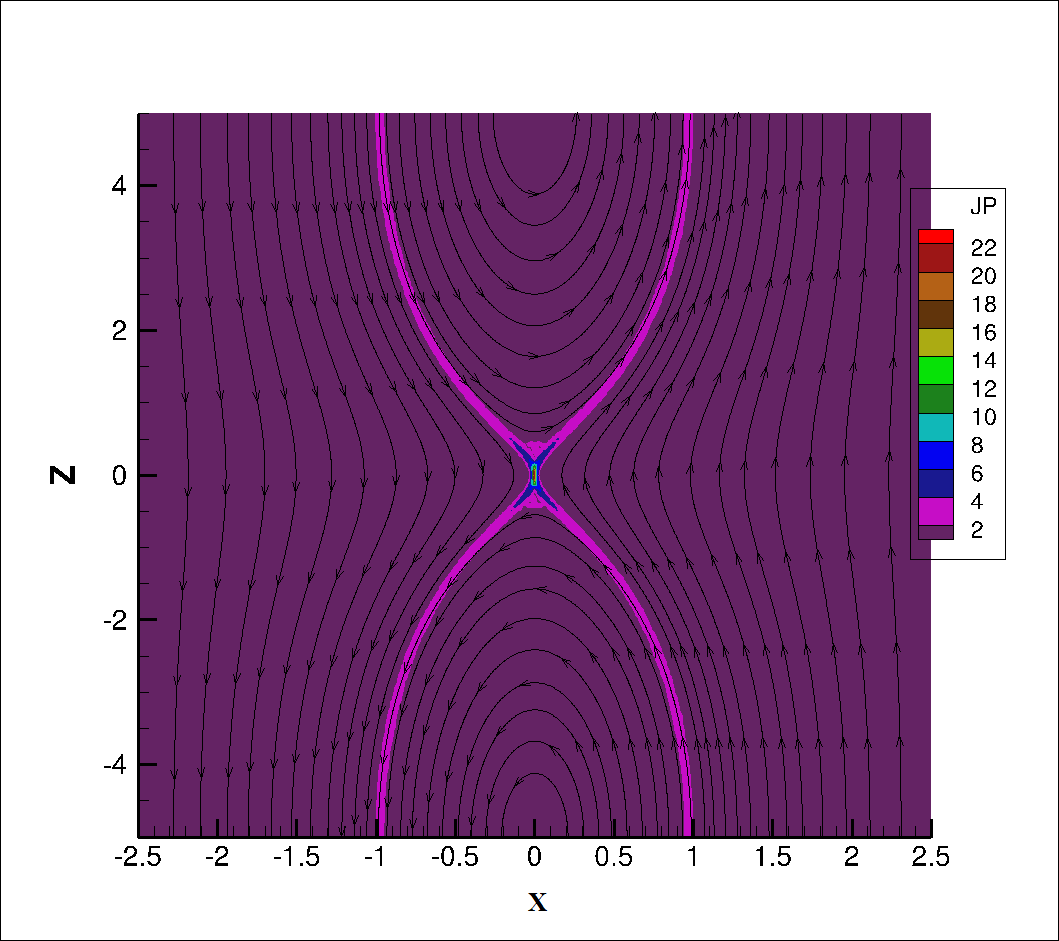}
\put(-250,225){\textbf{(b)}}
\end{minipage}
\caption{2D contours of the current density in z-direction and the 2D magnetic field lines at (a) time = 300 sec and (b) time = 470 sec for ${\Delta^\prime = 14.05}$.}
\end{figure}

\begin{figure}[htbp]
\includegraphics[width=1\linewidth]{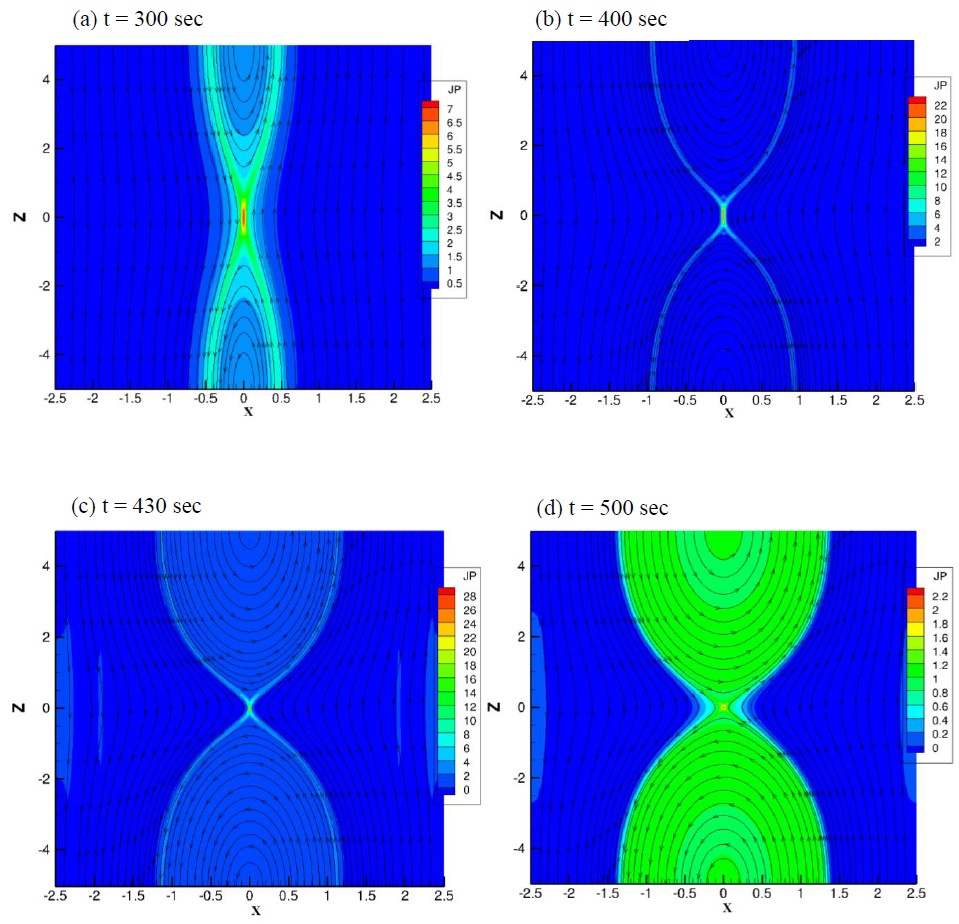}
\caption{2D contours of the current density in z-direction and the 2D magnetic field lines at (a) time = 300 sec, (b) time = 400 sec, (c) time = 430 sec and (d) time = 500 sec for ${\Delta^\prime = 15.6419}$.}
\label{fig:10}
\end{figure}

\begin{figure}[htbp]
\includegraphics[width=1\linewidth]{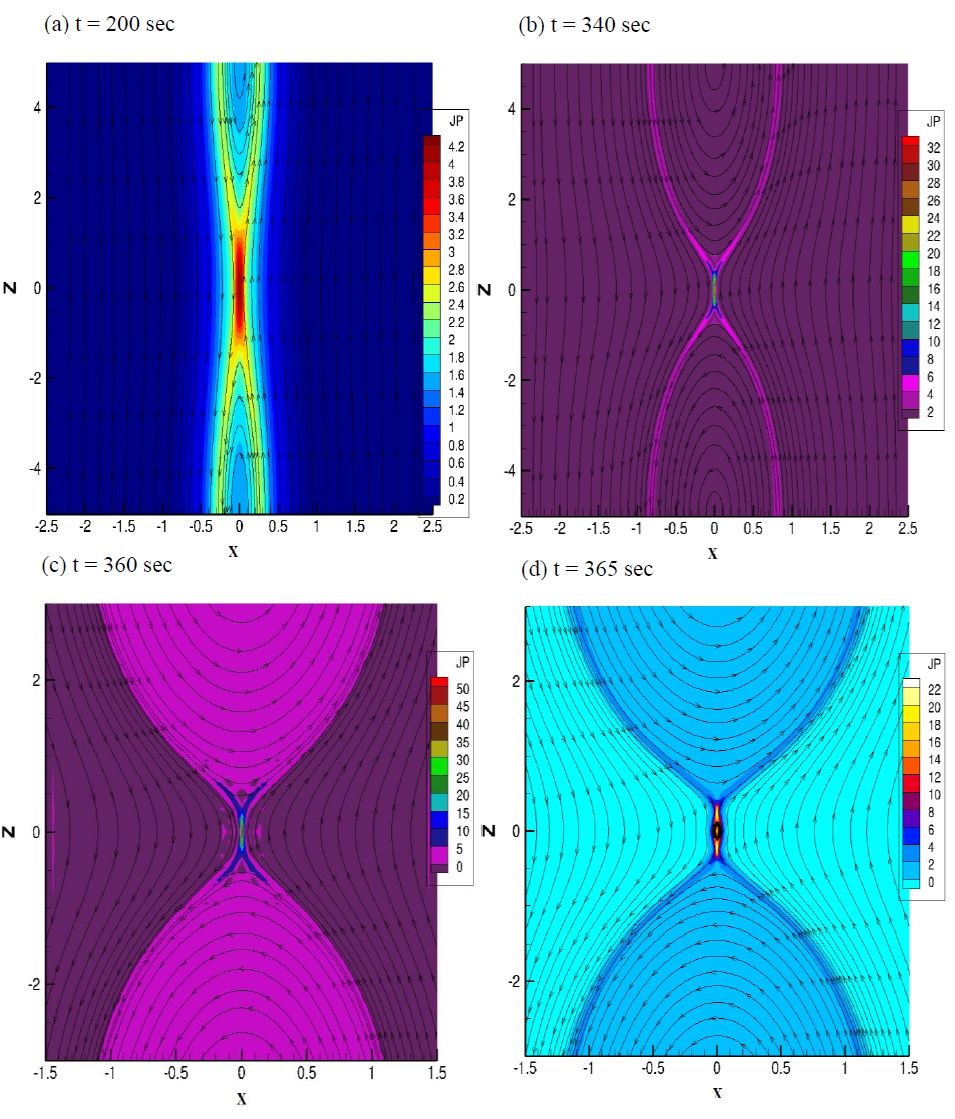}
\caption{2D contours of the current density in z-direction and the 2D magnetic field lines at  (a) time = 200 sec, (b) time = 340 sec, (c) time = 360 sec and (d) time = 365 sec for ${\Delta^\prime=17.31}$.}
\label{fig:10}
\end{figure}

\begin{figure}[htbp]
\includegraphics[width=1\linewidth]{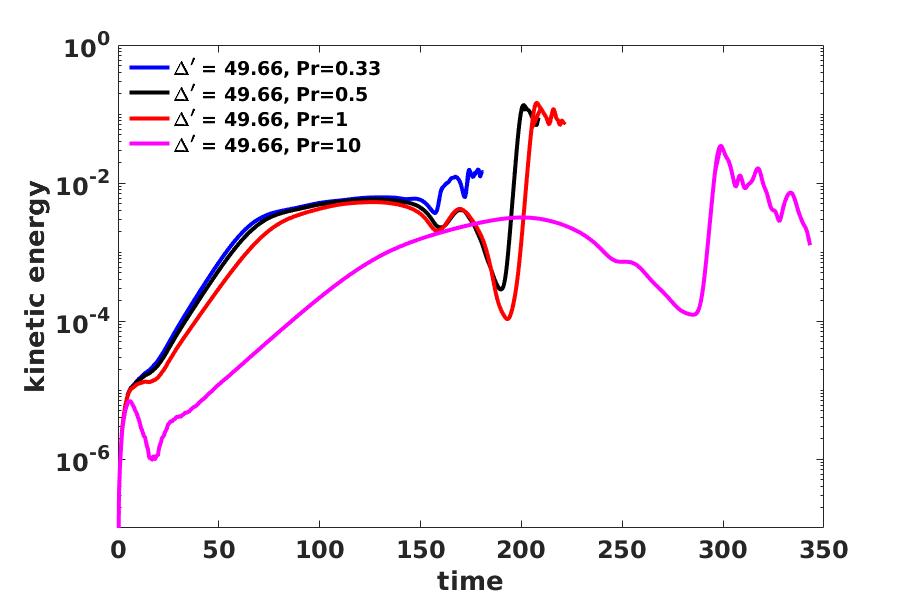}
\caption{Kinetic energy evolution for $\Delta^\prime$ = 49.66 at $P_r$ = 0.33, 0.5, 1 and $P_r$ = 10.}
\label{fig:10}
\end{figure}
\begin{figure}[htbp]
\includegraphics[width=1\linewidth]{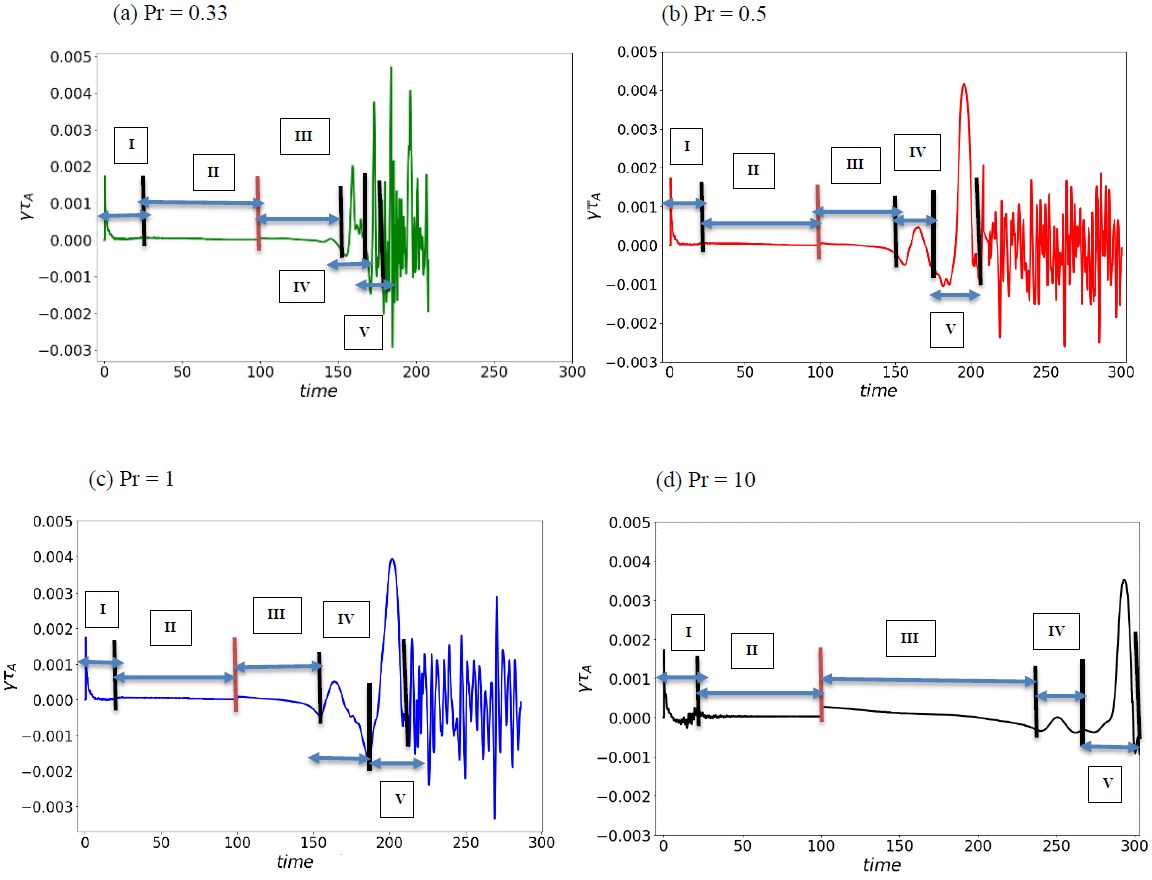}
\caption{Growth rate evolution for all ${P_r = 0.33, 0.5, 1}$ and ${10}$ cases with $\Delta^\prime$=49.66.}
\label{fig:10}
\end{figure}
\begin{figure}[htbp]
\includegraphics[width=1\linewidth]{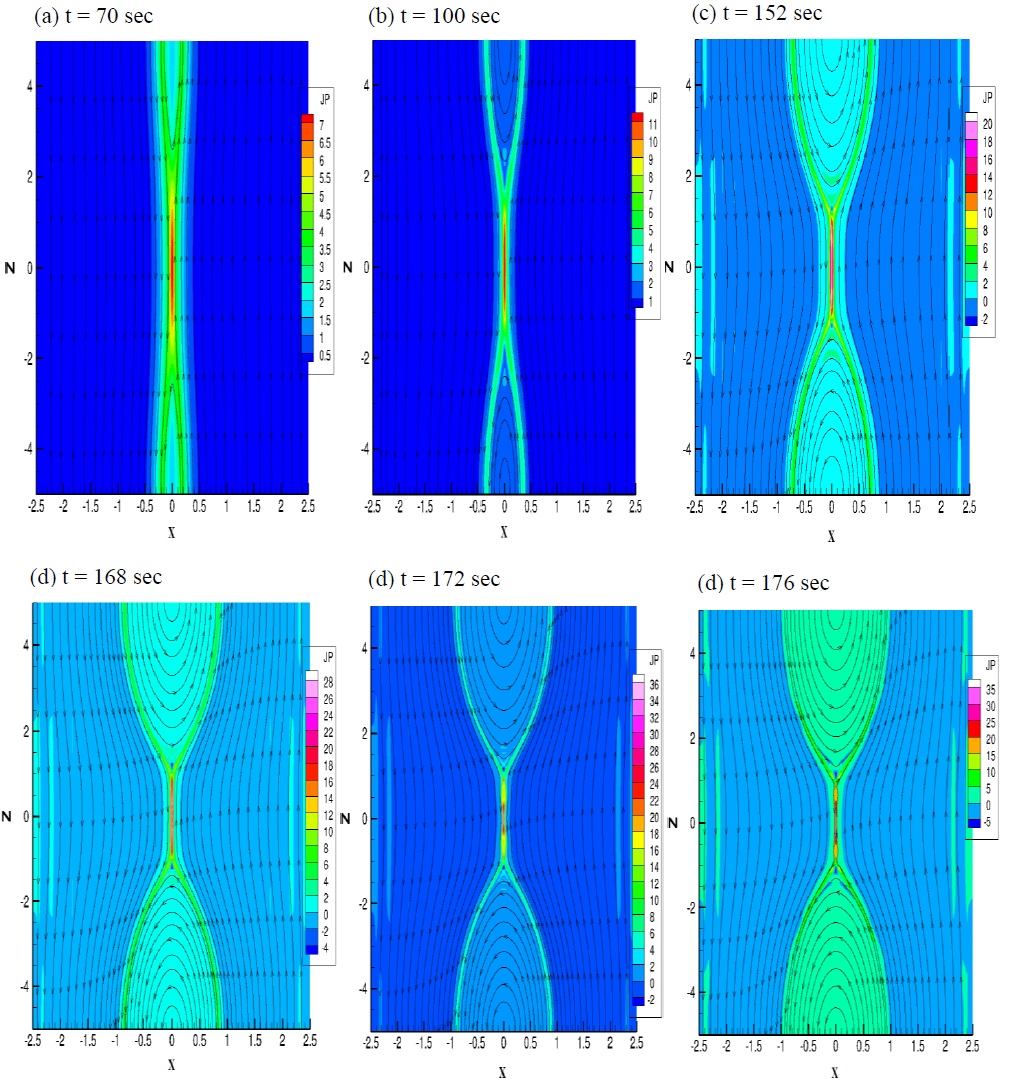}
\caption{2D contours of the current density in z-direction and the 2D magnetic field lines at (a) time = 70 sec, (b) time = 100 sec, (c) time = 152 sec, (d) time = 168 sec, (e) time = 172 sec and (f) time = 176 sec for {$\Delta^\prime$=49.66, ${P_r = 0.33}$}.}
\label{fig:10}
\end{figure}
\begin{figure}[htbp]
\includegraphics[width=1\linewidth]{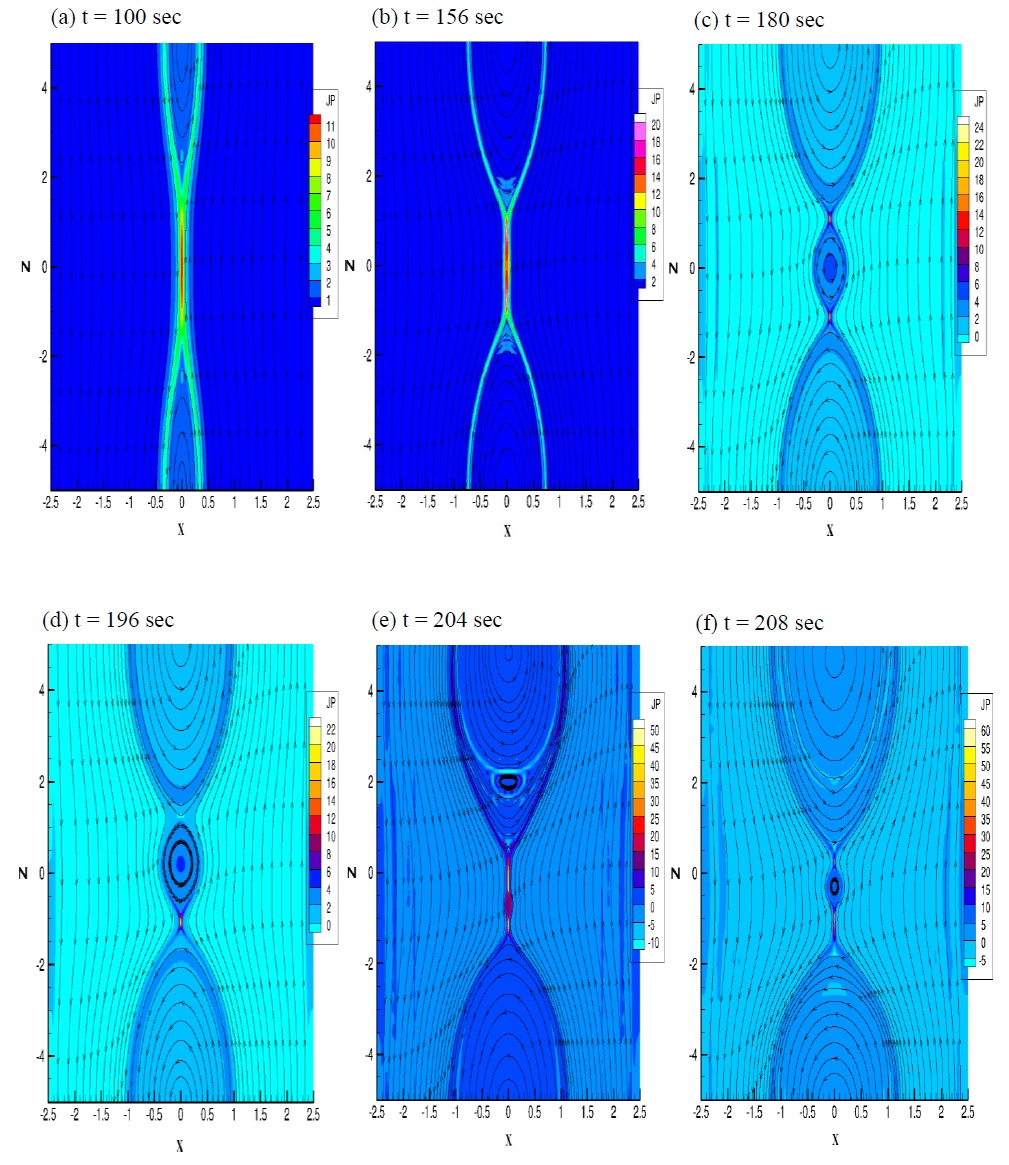}
\caption{2D contours of the current density in z-direction and the 2D magnetic field lines at (a) time = 100 sec, (b) time = 156 sec, (c) time = 180 sec, (d) time = 196 sec, (e) time = 204 sec and (f) time = 208 sec for {$\Delta^\prime$=49.66}, ${P_r = 0.5}$.}
\label{fig:10}
\end{figure}
\begin{figure}[htbp]
\includegraphics[width=1\linewidth]{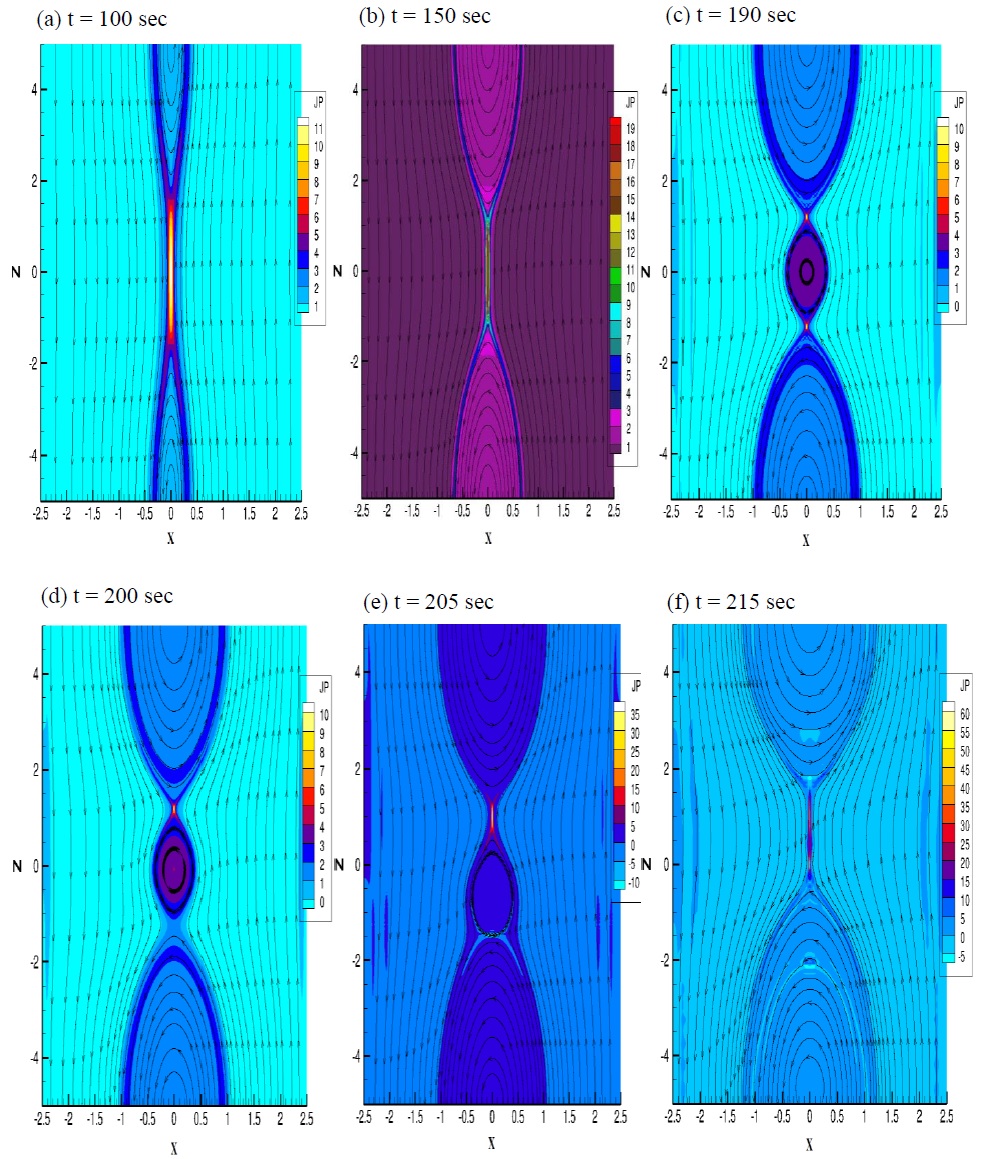}
\caption{2D contours of the current density in z-direction and the 2D magnetic field lines at (a) time = 100 sec, (b) time = 150 sec, (c) time = 190 sec, (d) time = 200 sec, (e) time = 205 sec and (f) time = 215 sec for {$\Delta^\prime$=49.66},  ${P_r = 1}$.}
\label{fig:10}
\end{figure}
\begin{figure}[htbp]
\includegraphics[width=1\linewidth]{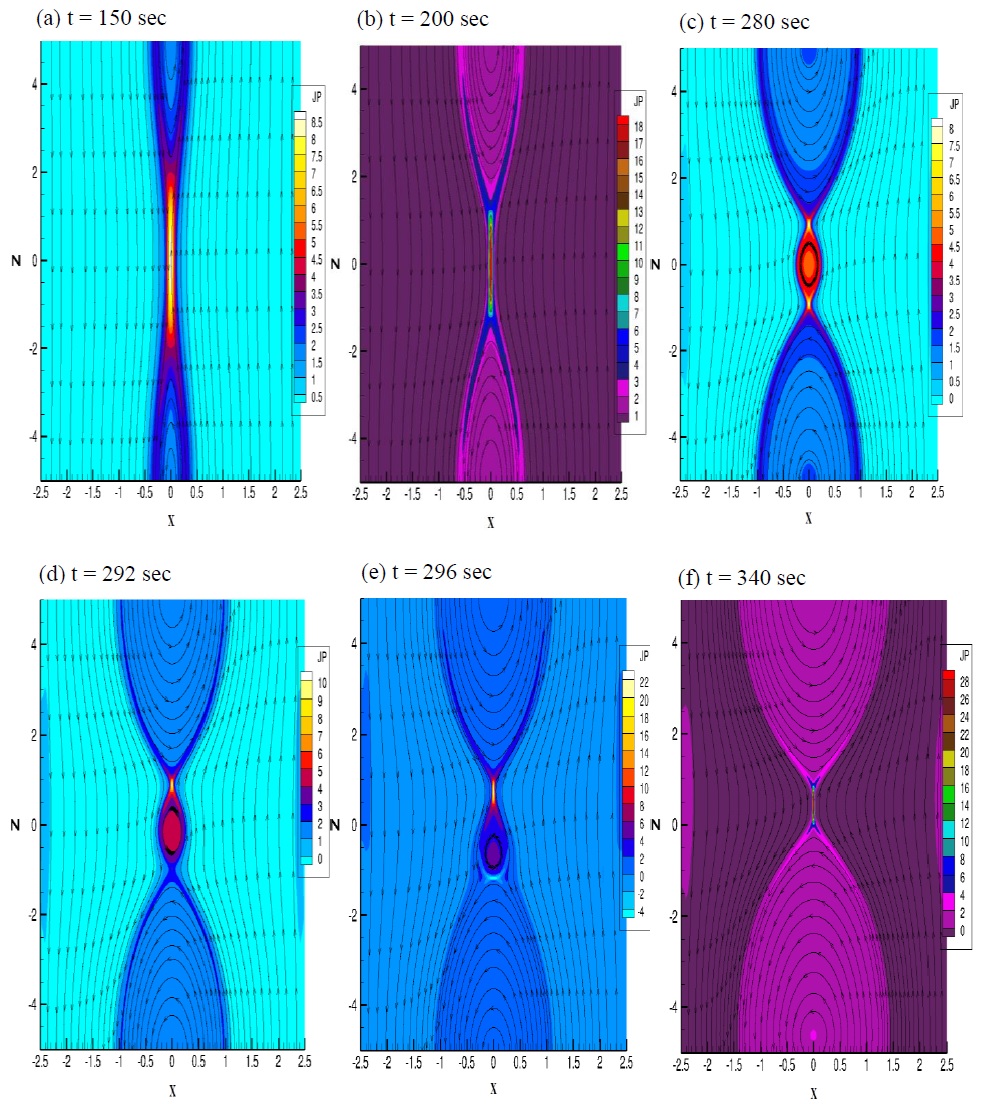}
\caption{2D contours of the current density in z-direction and the 2D magnetic field lines at (a) time = 150 sec, (b) time = 200 sec, (c) time = 280 sec, (d) time = 292 sec, (e) time = 296 sec and (f) time = 304 sec for {$\Delta^\prime$=49.66}, ${P_r = 10}$.}
\label{fig:10}
\end{figure}
\begin{figure}[htbp]
\centering
\begin{minipage}{1.0\textwidth}
\includegraphics[width=1.0\textwidth]{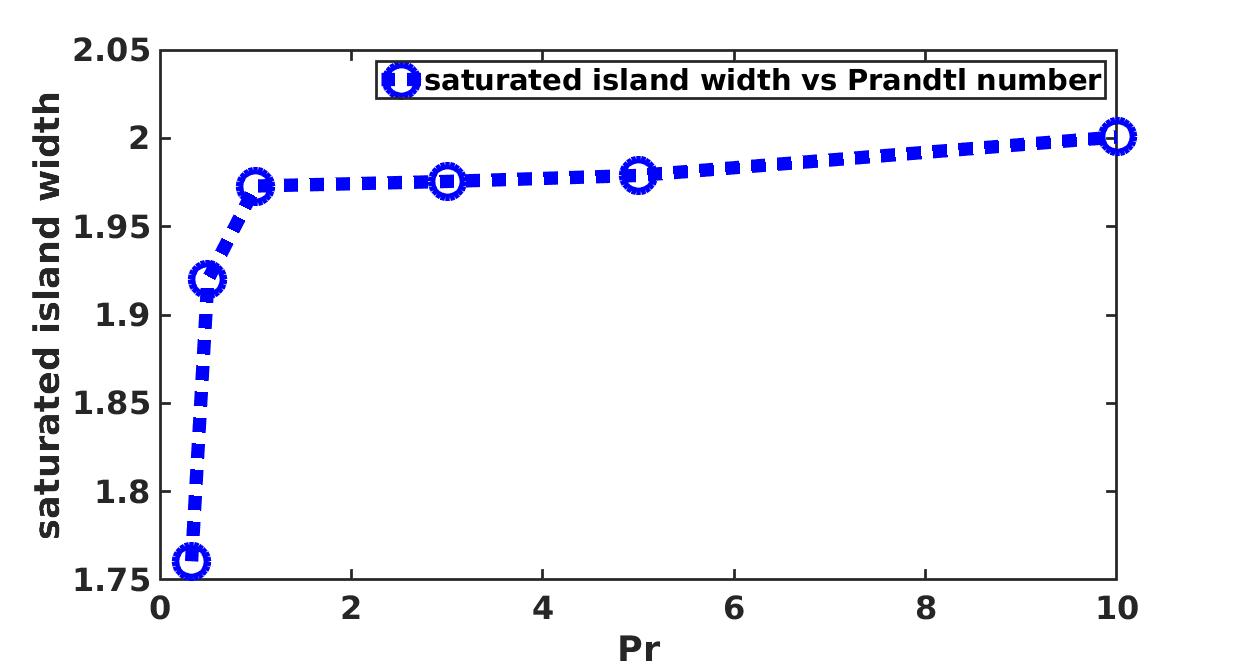}
\put(-390,230){\textbf{(a)}}
\end{minipage}
\begin{minipage}{1.0\textwidth}
\includegraphics[width=1.0\textwidth]{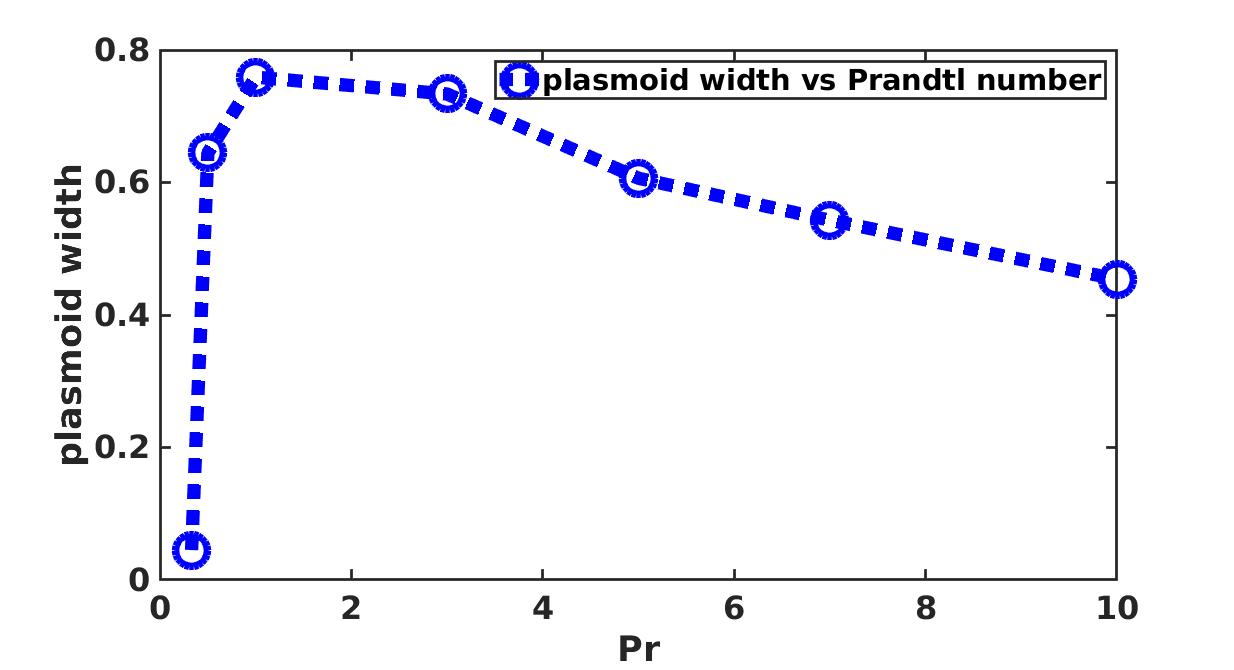}
\put(-390,230){\textbf{(b)}}
\end{minipage}
\caption{(a) Saturated island width and (b) plasmoid width as functions of the Prandtl number.}
\end{figure}

\subsubsection{Effect of viscosity on the non-linear evolution of resistive tearing mode for highly unstable system}

 ~\\In our cases, we only vary the viscosity in terms of ${P_r}$ by keeping the resistivity and instability parameter constant, which are ${\eta = 2.8\times10^{-4}}$ and ${\Delta^\prime = 49.66}$ respectively. To study the effect of viscosity, we choose four different cases with ${P_r = 0.33, 0.5,1}$ and ${10}$ (Figure 9). In Figure 10, the dynamics of the visco-resistive tearing mode growth are divided in 5-stages. The first stage is the initial transient stage when the linear instability starts to grow. The second stage is the ${FKR}$ stage, during which both the reconnection rate and the magnetic island width grow exponentially. In the third stage, the so-called Rutherford stage, the island evolves toward saturation and subsequent decay. The X-point collapse and the Y-type, SP-like current sheet forms during the fourth stage (Figure 11 (a)-(c)). The transition from the X-type geometry to the Y-type current sheet is known as the secondary instability. The first peak that appears in the kinetic energy evolution is due to the X-point collapse with the onset of secondary instability at t = 160 sec as shown in Figure 9. After the X-point collapse, the SP-like current sheet starts to become elongated in the poloidal direction. During the fifth stage, the Y-type current sheet starts to become more elongated and SP-like. After the collapse of SP-like current sheet, the secondary island appears along with two X-points on both ends (Figure 11(e)). After t = 160 sec, the growth rate starts to decrease up to t = 170 sec, and a significant change in the growth rate occurs due to the collapse of the Y-type current sheet and the formation of small plasmoid chain. The size of the secondary island increases up to some extent and both X-points collapse to form two tertiary current sheets with the passage of time. The second large peaks in the growth rate and the kinetic energy plots represent the collapse of these two tertiary current sheets. As the width of the secondary island approaches some critical value, the ejection of the secondary island takes place. At this point kinetic energy increases abruptly and the secondary island coalesces with the primary one (Figure 12(d)). As the Y-type current sheet collapse and the plasmoid instability (PI) appears, a drastic increase in the growth rate takes place (Figure 10) which is much larger as compared to the growth rate of the secondary instability.

 The nonlinear stages of our simulation results described above are quite similar to the nonlinear secondary island evolution reported by N. F. Loureiro et al. \cite{LOUREIRO2005}. But the collapse of the secondary island and direction of ejection are different in our cases. For example in our simulations for the ${P_r = 0.5}$ and the ${P_r = 5}$ cases, the direction of plasmoid ejection is upward, whereas for the ${P_r = 1}$ and the ${P_r = 10}$ cases the direction is downward. The directions of plasmoid ejection are different for different ${P_r}$ cases. 

 Our simulations for various ${P_r}$ numbers show viscous effects on both the timing and the spatial structure of plasmoid instability. In the ${P_r}$ = 0.33 (i.e. very low viscosity), we find secondary islands that saturate early at smaller size (Figure 11(f)). As we increase the viscosity further (${P_r}$ = 0.5), the appearance and ejection of the secondary island becomes more prominent, along with the larger saturated island and plasmoid (Figure 12). A clear transition occurs at ${P_r = 1}$, when the size of the primary plasmoid becomes the maximum, and the onsets of the secondary instability, the PI and the island saturation are significantly postponed. The direction of plasmoid ejection also switches to the opposite. The second peak of kinetic energy increases with viscosity and reaches maximum at ${P_r = 1}$ too. For higher Prandtl number (${P_r}$ = 10, with ${\eta = 2.8\times10^{-4}}$ and viscosity = 0.0028) the size of the plasmoid becomes smaller (Figure 14).

Figure 15 summarizes the relationships among plasmoid width, saturated island width, and ${P_r}$ number. At lower ${P_r}$, the width of saturated island is small, but as we increase viscosity, the width of saturated island increases sharply up to ${P_r = 1}$, beyond which the saturated island width becomes almost independent of the ${P_r}$ number. The ${P_r = 1}$ number also separates two regimes for the plasmoid width. In the ${P_r < 1}$ regime, the width of plasmoid increases drastically with viscosity, whereas in the ${P_r > 1}$ regime, the width of plasmoid slowly decreases with the viscosity.
\newpage
\section{Summary}

The key objective of this study is to explore the viscous effects on secondary instability, plasmoid formation, their merging and ejection process during the nonlinear evolution of a resistive tearing mode in the large ${\Delta^\prime}$ regime. For our equilibrium, we find the critical instability parameter for the onset of the secondary instability, and the minimum value of ${\Delta^\prime}$ at which PI can take place. Two distinctive regimes of the ${P_r}$ number are found for the plasmoid instability, which are separated by the value of ${P_r = 1}$. In the ${P_r \lesssim 1}$ regime, the amplitude of the second peak for kinetic energy increases up to ${P_r = 1}$, whereas in the ${P_r \gtrsim 1}$ regime, this amplitude decreases with the viscosity. Both the saturated island width and the plasmoid size increase sharply with the viscosity in the ${P_r \lesssim 1}$ regime, however, the former slowly increases whereas the later decreases with the viscosity in the ${P_r \gtrsim 1}$ regime. In another word, the plasmoid size reaches maximum at ${P_r \simeq 1}$. We plan to further explore the significance of such a finding in future work.  

\ack

This research was supported by the National Magnetic Confinement Fusion Science Program of China (No.2019YFE03050004), National Natural Science Foundation of China (Nos.11775221 and 51821005), U.S.DOE (Nos.DE-FG02-86ER53218 and DESC0018001), and the Fundamental Research Funds for the Central Universities at Huazhong University of Science and Technology (No.2019kfyXJJS193). We are grateful for the support from NIMROD team. This research used the computing resources from the Supercomputing Center of University of Science and Technology of China. The author Nisar Ahmad acknowledges the support from the Chinese Government Scholarship.

\newpage

\section*{References}
\providecommand{\newblock}{}


\end{CJK*}
\end{document}